\documentstyle[12pt,psfig]{article}

\newcommand{\resection}[1]{\setcounter{equation}{0}\section{#1}}

\newcommand{\EQ}{\begin{equation}}
\newcommand{\EN}{\end{equation}}
\newcommand{\bea}{\begin{eqnarray}}
\newcommand{\eea}{\end{eqnarray}}

\newcommand{\th}{\theta}

\begin{document}
\setcounter{page}{0}
\topmargin 0pt
\oddsidemargin 5mm
\renewcommand{\thefootnote}{\arabic{footnote}}
\newpage
\setcounter{page}{0}
\begin{titlepage}
\begin{flushright}
LPTHE/99-41
\end{flushright}
\vspace{0.5cm}
\begin{center}
{\large {\bf Correlators in integrable quantum field theory. The scaling RSOS 
models}}\\
\vspace{1.8cm}
{\large G. Delfino} \\ \vspace{0.5cm}
{\em Laboratoire de Physique Th\'eorique et Hautes Energies}\\
{\em Universit\'e Pierre et Marie Curie, Tour 16 $1^{er}$ \'etage, 4 place 
Jussieu}\\
{\em 75252 Paris cedex 05, France}\\
{\em E-mail: aldo@lpthe.jussieu.fr}\\
\end{center}
\vspace{1.2cm}

\renewcommand{\thefootnote}{\arabic{footnote}}
\setcounter{footnote}{0}

\begin{abstract}
\noindent
The study of the scaling limit of two-dimensional models of statistical 
mechanics within the framework of integrable field theory is illustrated 
through the example of the RSOS models. Starting from the exact description
of regime III in terms of colliding particles, we compute the correlation 
functions of the thermal, $\varphi_{1,2}$ and (for some cases) spin operators 
in the two-particle approximation. The accuracy obtained for the moments of 
these correlators is analysed by computing the central charge and the scaling
dimensions and comparing with the exact results. We further consider the 
(generally non-integrable) perturbation of the critical points with both the 
operators $\varphi_{1,3}$ and $\varphi_{1,2}$ and locate the branches solved 
on the lattice within the associated two-dimensional phase diagram. Finally we 
discuss the fact that the RSOS models, the dilute $q$-state Potts model at and 
the $O(n)$ vector model are all described by the same perturbed conformal field
theory.
\end{abstract}

\vspace{.3cm}

\end{titlepage}

\newpage
\resection{Introduction}

Integrable field theory emerged in the last years as an elegant and 
effective tool for the study of many two-dimensional statistical models 
directly in their scaling limit. The approach relies on the fact that a large
class of quantum field theories in (1+1) dimensions admits an infinite number
of integrals of motions (i.e. they are `integrable') and can be completely 
solved on-shell \cite{ZZ,Taniguchi}. The matrix elements of local operators on 
the asymptotic states are also exactly computable \cite{Karowski,Smirnov} and
lead to spectral series for the correlation functions whose quantitative 
effectiveness is remarkable. Among the statistical models that have been 
studied in this framework we mention the Ising model in a magnetic field
\cite{Taniguchi,immf,DS,ratios,nonint}, the $q$-state Potts model 
\cite{CZ,DC,DBC,dilute}, the $O(n)$ model \cite{Sashapolymers,CM}, 
and the Ashkin-Teller model \cite{AT}.

It is the purpose of this paper to illustrate how the programme outlined above
applies to the ``restricted solid-on-solid'' (RSOS) models introduced by 
Andrews, Baxter and Forrester in Ref.\,\cite{ABF}. For any integer $p\geq 3$ 
they are defined on the square lattice in terms of a spin or ``height'' 
variable $h_i$ located at each site $i$ and taking the integer values from $1$ 
to $p$. The heights of two nearest-neighbour sites $i$ and $j$ are constrained 
by the condition
\EQ
|h_i-h_j|=1\,\,,
\label{constraint}
\EN
which, in particular, leads to a natural splitting of the lattice into two
sublattices on which the heights are all even or all odd.
The Hamiltonian is further specified by a one-site and a diagonal interaction
terms. Their precise form does need to be reproduced here but it is 
important that the energy of a configuration is invariant under the global
tranformation
\EQ
h_i\rightarrow p+1-h_i\,\,,
\label{reflection}
\EN
which is the basic symmetry of the models.

In their general formulation the RSOS models contain a number of parameters
which grows linearly with $p$. In Ref.\,\cite{ABF} the models were solved
on two two-dimensional manifolds of the parameter space which can be 
parameterised
by a temperature-like variable $t$ together with a second coordinate $v$ 
measuring the spatial anisotropy of the lattice interaction. The scaling 
limit, however, is isotropic and $v$ will be ignored in the following.
Hence, for each of the two solutions and for any $p$, the phase diagram 
reduces to a line and exhibits a critical point separating two phases known as 
regimes I and II (for the first solution) and regimes III and IV (for the 
second solution). Here, we will only be interested in the second case, and 
more specifically in regime III.

By comparison of critical exponents, Huse showed \cite{Huse} that the critical
points separating regimes III and IV for the different values of $p$ 
correspond to the minimal unitary series of conformal field theories 
characterised by the values of the central charge \cite{BPZ,FQS}
\EQ
C=1-\frac{6}{p(p+1)}\,\,, \hspace{1cm}p=3,4,\ldots\,\,.
\label{c}
\EN
These models contain a finite number of primary operators $\varphi_{m,n}(x)$
($m=1,\ldots,p-1$; $n=1,\ldots,p$) with scaling dimensions
\EQ
X_{m,n}=\frac{[(p+1)m-pn]^2-1}{2p(p+1)}\,\,.
\label{xmn}
\EN

It was shown in Ref.\,\cite{LG} that the unitary minimal models admit the 
Landau-Ginzburg description 
\EQ
S=\int d^2x\,\left[(\partial\varphi)^2+\sum_{k=1}^{p-1}g_k\varphi^{2k}\right]
\,\,,
\label{lg}
\EN
with $g_1=g_2=\cdots=g_{p-2}=0$. The scalar field $\varphi(x)$ is then the 
continous version of the shifted height variable $h_i-(p+1)/2$, in such a
way that the reflection symmetry (\ref{reflection}) is mapped into 
$\varphi\rightarrow -\varphi$. The following identifications between normal 
ordered powers of $\varphi$ and conformal operators hold \cite{LG}
\EQ
\varphi^k\sim\varphi_{k+1,k+1}\,,\hspace{1cm}k=0,\ldots,p-2\,\,.
\label{powers}
\EN
The effective action (\ref{lg}) makes 
transparent that the considered series of critical points corresponds to 
the $(p-1)$-critical behaviour of a system with scalar order parameter and 
$Z_2$ symmetry, $p=3$ being the ordinary Ising universality class, $p=4$ the 
tricritical Ising, and so on. 

The RSOS models in regimes III and IV possess $p-1$ and $p-2$ degenerate 
ground states, respectively \cite{ABF}, and can be described by the
action (\ref{lg}) with suitable fine tunings of the couplings $g_k$ leading 
to the appropriate number of degenerate minima in the effective potential.
For $p=3$ the two regimes correspond to the low- and high-temperature phases
of the Ising model in zero magnetic field and are both massive in the scaling
limit. For $p\geq 4$, while regime III is still massive, regime IV become 
massless and corresponds to the crossover between the critical points labelled
by $p$ and $p-1$ \cite{Sashacth,CL}.

Solvability on the lattice naturally suggests integrability of the field theory
describing the scaling limit. In fact, the scaling dimension of the `thermal'
operator (conjugated to $t$) is known from the lattice solution and coincide
with $X_{1,3}$, so that the scaling limit of the RSOS models in 
regimes III and IV is described by the action
\EQ
{\cal A}={\cal A}_{CFT}^{(p)}+\lambda\int d^2x\,\varphi_{1,3}(x)\,\,,
\label{action}
\EN
where ${\cal A}_{CFT}^{(p)}$ is the action of the conformal theories with 
central charge (\ref{c}), and $\lambda$ is a coupling with dimensions 
$m^{2-X_{1,3}}$. This $\varphi_{1,3}$ perturbation of conformal field theory
is known to be massive (regime III) or massless (regime IV) depending on the 
sign of $\lambda$ \cite{Sashacth}, and to be integrable in both directions
\cite{Taniguchi}. The associated scattering theories are also known
\cite{SashaRSOS,LeClair,FSZ2}.

The paper is organised as follows. In the next section we briefly review the 
exact scattering description for regime III and use it in Section 3 for the 
computation of form factors of the operators $\varphi_{1,3}$, $\varphi_{1,2}$ 
and (for $p=3,4$) $\varphi_{2,2}$. In Section 4 we write down the two-particle 
approximation for the correlation functions of these operators and analyse its 
accuracy by computing the central charge and the scaling dimensions. 
Section 5 is devoted to a discussion of the perturbation of the RSOS 
critical points with both the operators $\varphi_{1,3}$ and $\varphi_{1,2}$. 
In the final section we briefly discuss 
the fact that the perturbed conformal field theory (\ref{action}) also 
describe the scaling limit of the dilute $q$-state Potts model along its 
first order phase transition lines for $0\leq q\leq 4$, as well as the $O(n)$ 
vector model for $-2\leq n \leq 2$.

\resection{Scattering theory}

In a (1+1)-dimensional theory with degenerate vacua the elementary excitations
are kinks interpolating among these vacua. It is known from the lattice
solution that the $j$-th ground state in regime III ($j=1,\ldots,p-1$) is such 
that all the sites on one sublattice have height $j$ and all the sites on the 
other sublattice have height $j+1$ (Fig. 1). The space-time trajectory of a 
kink is a domain wall separating two different ground states. Since the pairing
of two different ground states $i$ and $j$ can give an admissible configuration
only if $|i-j|$ equals $1$ (Fig. 2), the elementary excitations of the 
scattering theory are kinks\footnote{The rapidity variable $\th$ parameterises 
the on-shell momenta of the  kink of mass $m$ as $(p^0,p^1)=
(m\cosh\th,m\sinh\th)$.} $K_{ij}(\th)$ interpolating between two vacua $i$ and 
$j=i\pm 1$. It follows from the precise form of the lattice interaction that
the interfacial tension between two ground states $i$ and $i+1$ does not 
depend on $i$, and this amounts to say that the kinks $K_{i,i\pm 1}$ all have
the same mass. Multikink excitations of the type 
\EQ
\ldots K_{i\pm 1,i}(\th_1)K_{i,i\pm 1}(\th_2)\ldots
\EN
will connect ground states with arbitrary indices.

In an integrable field theory the scattering is completely elastic (no 
production processes allowed) and multiparticle processes factorise into
the product of the two-body subprocesses, so that the problem of the 
determination of the $S$-matrix is reduced to the computation of the two-kink 
amplitudes \cite{ZZ}.
Taking into account the kink composition rules together with invariance 
under time reversal and spatial inversion, the allowed two-kink processes 
are those depicted in Fig. 3 and associated to the commutation relations
\bea
K_{j,j\pm 1}(\th_1)K_{j\pm 1,j}(\th_2) &=& 
A^\pm_j(\th_1-\th_2)K_{j,j\pm 1}(\th_2)K_{j\pm 1,j}(\th_1)\\
&+& B_j(\th_1-\th_2)K_{j,j\mp 1}(\th_2)K_{j\mp 1,j}(\th_1)\,,\label{fz1}\\
K_{j\pm 1,j}(\th_1)K_{j,j\mp 1}(\th_2) &=& 
C_j(\th_1-\th_2)K_{j\pm 1,j}(\th_2)K_{j,j\mp 1}(\th_1)\,.
\label{fz2}
\eea
The scattering amplitudes are subject to a series of constraints.
Invariance under the reflection $j\rightarrow p-j$ requires
\bea
&& A^+_j(\th)=A^-_{p-j}(\th)\,,\\
&& B_j(\th)=B_{p-j}(\th)\,,\\
&& C_j(\th)=C_{p-j}(\th)\,,
\eea
while crossing symmetry implies
\bea
&& A^\pm_j(\th)=A^\mp_{j\pm 1}(i\pi-\th)\,,\\
&& B_j(\th)=C_j(i\pi-\th)\,\,.
\eea
Commuting once again the r.h.s. of Eqs.\,(\ref{fz1}) and (\ref{fz2}) leads 
to the unitarity equations
\bea
A^\pm_j(\th)A_j^\pm(-\th)+B_j(\th)B_j(-\th) &=& 1\,\,,\\
A^\pm_j(\th)B_j(-\th)+B_j(\th)A^\mp_j(-\th) &=& 0\,\,,\\
C_j(\th)C_j(-\th) &=& 1\,\,.
\eea
A three-kink process can be factorised in two ways differing by the ordering
of the two-body collisions. Equating the results leads to the factorisation
equation
\EQ
A^\pm_jA^\mp_{j\pm 1}A^\pm_j+B_jC_jB_j=
A^\mp_{j\pm 1}A^\pm_jA^\mp_{j\pm 1}+B_{j\pm 1}C_{j\pm 1}B_{j\pm 1}\,,
\EN
and similar others (the arguments of the three factors in each product are
$\th$, $\th+\th'$ and $\th'$, respectively).

The minimal solution to all these requirements is well known 
\cite{ABF,SashaRSOS,LeClair} and reads
\bea
&& A^\pm_j(\th)=\left(\frac{s_{j\pm 1}}{s_j}\right)^{i\th/\pi}
\frac{s_1}{s_j}\,\frac{\sinh\frac{1}{p}(ij\pi\pm\th)}
{\sinh\frac{1}{p}(i\pi-\th)}\,S_0(\th)\,\,,\\
&& B_j(\th)=\left(\frac{\sqrt{s_{j+1}s_{j-1}}}{s_j}\right)^{1+i\th/\pi}
\frac{\sinh\frac{\th}{p}}
{\sinh\frac{1}{p}(i\pi-\th)}\,S_0(\th)\,\,,\\
&& C_j(\th)=\left(\frac{\sqrt{s_{j+1}s_{j-1}}}{s_j}\right)^{-i\th/\pi}
\frac{\sinh\frac{1}{p}(i\pi-\th)}
{\sinh\frac{1}{p}(i\pi-\th)}\,S_0(\th)\,\,,
\eea
where 
\EQ
s_j\equiv \sin\frac{j\pi}{p}\,\,,
\EN
\bea
S_0(\th)=-\prod_{n=0}^\infty\frac{
\Gamma\left(1+\frac{2}{p}(n+\frac{1}{2})+\frac{\th}{i\pi p}\right)
\Gamma\left(1+\frac{2}{p}n-\frac{\th}{i\pi p}\right)}{
\Gamma\left(1+\frac{2}{p}(n+\frac{1}{2})-\frac{\th}{i\pi p}\right)
\Gamma\left(1+\frac{2}{p}n+\frac{\th}{i\pi p}\right)}&&\nonumber \\
         \times\frac{
\Gamma\left(\frac{2}{p}(n+1)-\frac{\th}{i\pi p}\right)
\Gamma\left(\frac{2}{p}(n+\frac{1}{2})+\frac{\th}{i\pi p}\right)}{
\Gamma\left(\frac{2}{p}(n+1)+\frac{\th}{i\pi p}\right)
\Gamma\left(\frac{2}{p}(n+\frac{1}{2})-\frac{\th}{i\pi p}\right)}&&\nonumber \\
        =-\exp\left\{i\int_0^\infty\frac{dx}{x}\,
\frac{\sinh(p-1)\frac{x}{2}}{\sinh\frac{px}{2}\cosh\frac{x}{2}}\,
\sin\frac{x\th}{\pi}\right\}\,\,.\hspace{1cm}&&
\eea
It can be checked that the amplitudes do not posses poles in the physical 
strip $\mbox{Im}\th\in(0,\pi)$, what ensures that there are no bound states 
and that the amplitudes given above entirely determine the scattering theory.

\resection{Form factors}

Let us denote $\Phi(x)$ a local scalar operator of the theory with zero 
topological charge, namely such that its action on the vacuum $|0_j\rangle$ 
only produces excitations beginning and ending on this vacuum. All the
operators we will consider in the following share this property. We are 
interested in the two-particle form factors (Fig. 4)
\EQ
F^\Phi_{j,\pm}(\th_1-\th_2)=\langle 0_j|\Phi(0)|K_{j,j\pm 1}(\th_1)
K_{j\pm 1,j}(\th_2)\rangle\,\,.
\EN
Eq.\,(\ref{fz1}) implies the relation
\EQ
F^\Phi_{j,\pm}(\th)=A^\pm_j(\th)F^\Phi_{j,\pm}(-\th)+B_j(\th)
F^\Phi_{j,\mp}(-\th)\,\,,
\label{ffunitarity}
\EN
while crossing leads to the equations \cite{DC}
\EQ
F^\Phi_{j,\pm}(\th+2i\pi)=F^\Phi_{j\pm 1,\mp}(-\th)\,\,,
\label{ffcrossing}
\EN
\EQ
-i\,\mbox{Res}_{\th=i\pi}F^\Phi_{j,\pm}(\th)=
 i\,\mbox{Res}_{\th=i\pi}F^\Phi_{j\pm 1,\mp}(\th)=
\langle 0_j|\Phi|0_j\rangle - \langle 0_{j\pm 1}|\Phi|0_{j\pm 1}\rangle\,\,.
\label{annihilation}
\EN
As a last necessary condition, the two-kink form factors are subject to the 
asymptotic bound \cite{immf}
\EQ
\lim_{\th\rightarrow +\infty}F^\Phi_{j,\pm}(\th)\leq\mbox{constant}\,\,
e^{X_\Phi\th/2}\,\,,
\label{bound}
\EN
where $X_\Phi$ denotes the scaling dimension of the operator $\Phi(x)$.

It is easily checked that a class of solutions of Eqs.\,(\ref{ffunitarity}) 
and (\ref{ffcrossing}) is given by
\EQ
F^\phi_{j,\pm}=-\frac{2i}{p}\,\left(\frac{s_{j\pm 1}}{s_j}\right)^{(1+
i\th/\pi)/2}\frac{F_0(\th)}{\sinh\frac{1}{p}(\th-i\pi)}\,
\Omega^\phi_j(\th)\,,
\label{subset}
\EN
where $\{\phi\}\subset\{\Phi\}$,
\EQ
F_0(\th)=-i\sinh\frac{\th}{2}\,\exp\left\{\int_0^\infty\frac{dx}{x}\,
\frac{\sinh(1-p)\frac{x}{2}}{\sinh\frac{px}{2}\cosh\frac{x}{2}}\,
\frac{\sin^2(i\pi-\th)\frac{x}{2\pi}}{\sinh x}\right\}
\EN
is solution of the equations
\bea
&& F_0(\th)=S_0(\th)F_0(-\th)\,,\\
&& F_0(\th+2i\pi)=F_0(-\th)\,,
\eea
and behave as 
\EQ
F_0(\th)\sim\exp[(1+1/p)\th/4]\,,\hspace{1cm}
\th\rightarrow+\infty\,. 
\EN
The functions $\Omega^\phi_j(\th)$ are free of 
poles and satisfy
\bea
&& \Omega^\phi_j(\th)=\Omega^\phi_j(-\th)\,,\\
&& \Omega^\phi_j(\th+2i\pi)=-\Omega^\phi_{j\pm 1}(-\th)\,\,.
\eea
These requirements, together with (\ref{bound}) imply that the
$\Omega^\phi_j(\th)$ are polynomials in $\cosh(\th/2)$. Let us consider 
those operators $\phi(x)$ which are relevant in the renormalisation group
sense ($X_\phi<2$) for all values of $p\geq 3$. Then, the bound (\ref{bound})
implies that these polynomials are at most of degree one, which means that
the operator subspace $\{\phi\}$ contains only two independent relevant 
operators. The trace of the 
stress-energy tensor $\Theta(x)$ is a relevant operator and is even under the
reflection symmetry, so that $F^\Theta_{j,+}(\th)=F^\Theta_{p-j,-}(\th)$. 
Moreover $\langle 0_j|\Theta|0_j\rangle$ does not depend on $j$ and, 
according to (\ref{annihilation}), the two-kink matrix elements have no pole
in $\th=i\pi$. All these requirements are fulfilled if we take
\EQ
\Omega^\Theta_j(\th)=2\pi m^2 \cosh\frac{\th}{2}\,\,,
\label{theta}
\EN
with the normalisation constant fixed by 
\EQ
F^\Theta_{j,\pm}(i\pi)=2\pi m^2\,\,.
\label{norm}
\EN
The other independent relevant operator in $\{\phi\}$ (let us denote it 
${\cal E}(x)$) corresponds to the constant solution
\EQ
\Omega^{\cal E}_j(\th)=(-1)^j\langle 0_1|{\cal E}|0_1\rangle\,\,.
\label{12}
\EN

The order parameter $\varphi(x)$ is the most relevant operator which changes 
sign under reflection, and this means in particular
\EQ
\langle 0_j|\varphi|0_j\rangle=-\langle 0_{p-j}|\varphi|0_{p-j}\rangle\,\,,
\EN
\EQ
F^\varphi_{j,+}(\th)=-F^\varphi_{p-j,-}(\th)\,\,.
\EN
For generic values of $p$, these properties are incompatible with those of 
the space of solutions spanned by (\ref{subset}), (\ref{theta}), (\ref{12})
and we conclude that $\varphi\not\in\{\phi\}$. We do not dispose of the
solution for $F^\varphi_{j,+}(\th)$ for generic $p$ and just quote the 
result for the two simplest cases
\EQ
F^\varphi_{1,+}(\th)=i\,\langle 0_1|\varphi|0_1\rangle\tanh\frac{\th}{2}\,,
\hspace{1cm}p=3
\EN
\EQ
F^\varphi_{j,+}(\th)=\frac{(-1)^j}{2\Upsilon_j(i\pi)}\,
\langle 0_1|\varphi|0_1\rangle
\left(\frac{s_{j+1}}{s_j}\right)^{(1+i\th/\pi)/2}\frac{F_0(\th)}{\cosh
\frac{\th}{2}}\,\Upsilon_j(\th)\,,\hspace{1cm}p=4\,.
\EN
The functions
\EQ
\Upsilon_j(\th)=\exp\left\{\int_0^\infty\frac{dx}{x}\,
\frac{\sin^2[2i\pi|2-j|-\th]\frac{x}{2\pi}}{\cosh x\,\sinh 2x}\right\}
\EN
satisfy the equations
\bea
&& \Upsilon_1(\th)=\Upsilon_3(\th)=\frac{\sinh\frac{1}{4}(i\pi+\th)}
{\sinh\frac{1}{4}(i\pi-\th)}\,\Upsilon_1(-\th)\,,\\
&& \Upsilon_1(\th+2i\pi)=\Upsilon_2(\th)\,,
\eea
and behave as $\exp(\th/8)$ when $\th\rightarrow +\infty$.
It can be checked that $F^\varphi_{j,\pm}(\th)=F^{\cal E}_{j,\pm}(\th)$ 
for $p=3$.

\resection{Correlation functions}

The correlation functions are obtained using the resolution of the identity
\EQ
1=\sum_{n=0}^\infty\int_{\th_1>\ldots>\th_n}\frac{d\th_1\ldots d\th_n}
{(2\pi)^n}|n\rangle\langle n|
\EN
to sum over all intermediate $n$-kink states $|n\rangle$. A two-point function 
reads\footnote{Here and in the following we always refer to connected 
correlators.}
\bea
\langle 0_j|\Phi_1(x)\Phi_2(0)|0_j\rangle &=& \sum_{\varepsilon=\pm}
\int_{\th_1>\th_2}\frac{d\th_1}{2\pi}\frac{d\th_2}{2\pi}
F^{\Phi_1}_{j,\varepsilon}(\th_1-\th_2)F^{\Phi_2}_{j,\varepsilon}
(\th_2-\th_1)\,e^{-|x|E_2}\\
&+& O(e^{-4m|x|})\,,\hspace{1cm}m|x|\gg 1
\label{approx}
\eea
where $E_2=m(\cosh\th_1+\cosh\th_2)$ is the energy of the two-kink asymptotic 
state. The ``two-kink approximation'' (\ref{approx}) is known to provide
results of remarkable accuracy for integrated correlators (see \cite{DC} and 
references therein), as can be checked through the use of the sum rules 
\cite{Cardycth,DSC}
\bea
&& C=\frac{3}{4\pi}\int d^2x\,|x|^2\langle 0_j|\Theta(x)\Theta(0)|
0_j\rangle\,,\label{sumc}\\
&& X_\Phi=-\frac{1}{2\pi\langle 0_j|\Phi|0_j\rangle}\,
\int d^2x\,\langle 0_j|\Theta(x)\Phi(0)|0_j\rangle\,,
\label{sumx}
\eea
allowing the determination of the ultraviolet conformal data (central charge 
and scaling dimensions) in the form of moments of off-critical correlators. 

In Fig. 5 we compare the exact formula for the central charge (\ref{c})
with the result yielded by the two-kink approximation (\ref{approx}) in the 
sum rule (\ref{sumc}). Notice that $p$ can be considered as a continous 
parameter in the result of the latter computation, in agreement with the fact 
that many
observables in the theory (\ref{action}) have a continous $p$-dependence.
With this remark in mind, in the remaining part of this section we will treat 
$p$ as a real number\footnote{In doing that, of course we loose unitarity 
unless $p$ is an integer larger than $2$. Here and in the following the term 
`unitarity' refers to the absence of negative norm states in the Hilbert 
space.} $\geq 1$. 

For $p=1$, in particular, the two-kink form factor (\ref{subset}), 
(\ref{theta}) simply reduces to $2\pi m^2(-1)^{(1+i\th/\pi)/2}$. This 
residual rapidity dependence ensures the normalisation condition (\ref{norm})
but is immaterial in the computation of the correlator (\ref{approx}). Hence,
the theory (\ref{action}) at $p=1$ is free and the two-kink computation for 
the central charge gives the exact result $C=-2$. Due to the equivalence  
between the Ising model and a free neutral fermion, the two-kink approximation
is exact also for $p=3$ and gives $C=-1/2$.

The computation of $X_\Theta$ through the sum rule (\ref{sumx}) requires the
knowledge of $\langle\Theta\rangle$. Although this quantity cannot be 
related to $F^\Theta_{j,\pm}(\th)$ due to the vanishing of the residue
(\ref{annihilation}), we dispose of the thermodynamic Bethe ansatz result
\cite{FSZ2}
\EQ
\langle\Theta\rangle=-\pi m^2\tan\frac{\pi p}{2}\,\,.
\EN
Since $\varphi_{1,3}(x)$ is the operator which drives the theory away from 
criticality, we must have 
\EQ
\Theta(x)\sim\lambda\varphi_{1,3}(x)\,\,,
\EN
and $X_\Theta=X_{1,3}$. 

The two-kink computation of $X_{\cal E}$ through (\ref{sumx}) gives $-1/4$ 
at $p=1$ and $1/8$ at $p=3$. Since the theory is free for these two values
of $p$, these results are expected to be exact. Assuming that ${\cal E}(x)$ 
corresponds to a primary operator whose position in the Kac's table does not 
depend on $p$, they can be substituted into the formula (\ref{xmn}) to fix 
\EQ
{\cal E}(x)\sim\varphi_{1,2}(x)\,\,.
\EN 
In Fig. 6 we compare the two-kink approximation for $X_\Theta$ and 
$X_{\cal E}$ with the exact formulae 
\bea
&& X_{1,3}=2\,\frac{p-1}{p+1}\,\,,\label{x13}\\
&& X_{1,2}=\frac{p-2}{2\,(p+1)}\label{x12}\,\,.
\eea
respectively.
Concerning the two values of $p$ for which we determined the form factors
of the order parameter, the two-kink computation gives $X_\varphi=1/8$ at
$p=3$ and $X_\varphi=0.0734$ at $p=4$. In view of the identification
$\varphi(x)\sim\varphi_{2,2}(x)$, these results must be 
compared with the exact values $1/8$ and $3/40=0.075$, respectively.

Some comments are in order about the results yielded by the sum rules
(\ref{sumc}) and (\ref{sumx}). 
Consider the moment 
\EQ
I_k=\int d^2x|x|^k\langle\Phi_1(x)\Phi_2(0)\rangle\,\,,
\label{moment}
\EN
and denote $\Phi_3(x)$ the leading operator determining the short distance
behaviour
\EQ
\langle\Phi_1(x)\Phi_2(0)\rangle\sim\frac{\langle\Phi_3\rangle}{|x|^{\eta_{
\Phi_1\Phi_2}}}\,,\hspace{1cm}|x|\rightarrow 0
\label{ope}
\EN
with
\EQ
\eta_{\Phi_1\Phi_2}\equiv X_{\Phi_1}+X_{\Phi_2}-X_{\Phi_3}\,\,.
\label{eta}
\EN
Then $I_k$ is convergent if $2+k-\eta_{\Phi_1\Phi_2}>0$.
In the cases of interest for us we have
$\eta_{\Theta\Theta}=2X_{1,3}$ and $\eta_{\Theta{\cal E}}=X_{1,3}$, so 
that the sum rules for $C$ and $X_{1,2}$ converge for all finite values of $p$,
while the sum rule for $X_{1,3}$ converges only for $p<3$. 
This `failure' of the sum rule for the scaling dimension due to 
the divergence of the integral is originated by operator mixing under 
renormalisation and was discussed in Ref.\,\cite{DSC}.

Let us now discuss the issue of the accuracy of the results obtained using 
the approximated correlators (\ref{approx}) into the integrals (\ref{sumc})
and (\ref{sumx}). The spectral series for the correlation functions is a 
large distance expansion and any partial sum including the contributions up 
to $n$ particles will appreciably depart from the exact result at sufficiently 
small distances. In the moment (\ref{moment}), however, the factor $|x|^k$ 
causes a suppression of the short distance contribution whose importance, for
a fixed $k$, depends on the high energy behaviour of the form factors.
The two-kink contribution to $I_k$ is given by
\EQ
\int d\th\,\frac{F^{\Phi_1}_{j,\pm}(2\th)F^{\Phi_2}_{j,\pm}(-2\th)}
{(\cosh\th)^{k+2}}\,\,.
\label{twop}
\EN
The integrand here behaves asymptotically as $\exp[-\Sigma_{\Phi_1\Phi_2}^{(k)}
\th]$, where
\EQ
\Sigma_{\Phi_1\Phi_2}^{(k)}\equiv 2+k-y_{\Phi_1}-y_{\Phi_2}\,\,,
\label{sigma}
\EN
with $y_\Phi$ defined by
\EQ
F^\Phi_{j,\pm}(\th)\sim e^{y_\Phi\th/2}\,\,,\hspace{1cm}\th\rightarrow
+\infty\,\,.
\EN
In an unitary theory, this exponent is subject to the bound \cite{immf}
\EQ
y_\Phi\leq X_\Phi\,\,.
\label{yx}
\EN
We see that $\Sigma_{\Phi_1\Phi_2}^{(k)}$ has to be positive to ensure the 
convergence of the integral and that the suppression of the short distance 
contribution is proportional to this exponent. Of course, this observation does
not determine the absolute accuracy of the two-particle approximation, but 
it helps understanding the accuracy pattern exhibited 
in Figs. 6. In fact, the solutions of Section 3 determine
\bea
&& y_\Theta=\frac{3(p-1)}{2p}\,\,,\\
&& y_{\cal E}=\frac{p-3}{2p}\,\,.
\eea
Then, $\Sigma^{(0)}_{\Theta\Theta}$ goes to zero as $p\rightarrow 3$ and we
observe that the deviation of the two-kink approximation for $X_\Theta$ from
the exact result becomes large as we approach this value. Analogous 
considerations apply to the case of $X_{\cal E}$ as $p\rightarrow\infty$. 
In the central charge sum rule (\ref{sumc}), on the contrary, 
$\Sigma^{(2)}_{\Theta\Theta}$ tends to 1 as $p\rightarrow\infty$, what means
that the high energy contribution is still strongly suppressed in this limit.
The remarkable accuracy (1\%) of the two-kink approximation
as $p\rightarrow\infty$ shows that also the contributions with a larger number
of kinks undergo a similar suppression . The relation between 
$\Sigma_{\Phi_1\Phi_2}^{(k)}$
and the accuracy of the two-kink approximation (defined as the absolute
deviation from the exact result divided by the exact result) is illustrated
in Table 1 through few examples.

\resection{Double perturbation and phase diagram}

In this section we briefly consider what happens if we add to the action
(\ref{action}) the operator ${\cal E}\sim\varphi_{1,2}$, namely if we 
take\footnote{When useful in this section we explicitely label the operators 
by the superscript $(p)$ identifying the critical point they refer to.}
\EQ
{\cal A}'={\cal A}^{(p)}_{CFT}+\lambda\int d^2x\,\varphi^{(p)}_{1,3}(x)+
\mu\int d^2x\,\varphi^{(p)}_{1,2}(x)\,\,.
\label{double}
\EN
Within the usual conventions for the operator normalisations, the regime III
we considered in the previous sections corresponds to $\mu=0$ and $\lambda<0$.
If $\mu$ is very small we can use form factor perturbation theory \cite{nonint}
around regime III. The correction to the energy density $\varepsilon_j$ of 
the vacuum $|0_j\rangle$ is proportional to the vacuum expectation value of the
perturbing operator $\varphi_{1,2}$ computed at $\mu=0$, and reads (remember
(\ref{12}) and (\ref{annihilation}))
\EQ
\delta\varepsilon_j\sim\mu\,\langle 0_j|\varphi_{1,2}|0_j\rangle=
\mu\,(-1)^j\langle 0_1|\varphi_{1,2}|0_1\rangle\,\,.
\EN
This means that only a subset of alternating vacua among the $p-1$ degenerate 
vacua of regime III preserves the same energy when $\lambda$ is negative and 
$\mu$ is small in (\ref{double}). For $p$ odd, in particular, we see that the
$Z_2$ symmetry characteristic of the case $\mu=0$ is broken by the 
$\varphi_{1,2}$ perturbation. For $p=3$ the action (\ref{double}) describes
the Ising model in a magnetic field \cite{ratios,nonint}. If $p$ is odd the 
number of surviving degenerate vacua is $(p-1)/2$. If $p$ is even, instead, 
this number is $p/2$ or $p/2-1$ depending on the sign of $\mu$. 

It is clear that in presence of such a pattern of degeneracy breaking the 
kinks $K_{j,j\pm 1}$ of regime III are no longer asymptotic excitations when
$\mu\neq 0$. Rather, they will be confined into pairs $K_{j,j+1}K_{j+1,j+2}$ 
providing the new stable kinks in the perturbed theory. This phenomenon
appears in the formalism when we try to compute the correction to the mass
of the kinks $K_{j,j\pm 1}$, which is given by \cite{nonint,DM}
\EQ
\delta m\sim\mu\,F^{\varphi_{1,2}}_{j,\pm}(i\pi)\,\,.
\EN
Since the form factor on the r.h.s. has a pole at $\th=i\pi$ (see 
(\ref{subset}) and (\ref{12})), it follows that this correction is infinite,
a fact that reveals the removal of the kinks $K_{j,j\pm 1}$ from the spectrum
of the asymptotic excitations.

The conformal field theories with $C<1$ perturbed with {\em one} of the 
operators $\varphi_{1,3}$, $\varphi_{1,2}$ or $\varphi_{2,1}$ are integrable 
\cite{Taniguchi}. It is then natural to look for solvable lattice models 
whose scaling limit corresponds to these quantum field theories.
In Ref.\,\cite{WPSN} a `dilute' version of the RSOS models was considered and 
found to be solvable on the lattice along four distinct branches. It was 
found, in particular, that the scaling limit of `branch 2' is described by the
action (\ref{double}) with $\lambda=0$, and that for $p$ odd this branch
possesses $(p-1)/2$ degenerate ground states. This result is consistent with
our perturbative considerations and suggests that they hold true for the whole
region $\lambda\leq 0$ in (\ref{double}). 

We already mentioned that for $p>3$ the regime IV of the RSOS models ($\lambda>
0$, $\mu=0$ in (\ref{double})) corresponds to a massless flow to the critical 
point with action ${\cal A}^{(p-1)}_{CFT}$. It is known that the operator
$\varphi_{1,2}^{(p)}$ renormalises into the operator $\varphi_{2,1}^{(p-1)}$
in the infrared limit of this flow \cite{Sashacth}. Hence we conclude that
the action (\ref{double}) in the limit $\lambda=+\infty$ describes the 
$\varphi_{2,1}^{(p-1)}$ perturbation of the critical point ${\cal A}^{(p-1)}_
{CFT}$. This integrable perturbation was identified in \cite{WPSN} as
corresponding to the scaling limit of the lattice models along `branch 
1'\footnote{The remaining two solvable branches of Ref.\,\cite{WPSN} are not
related to perturbations of the critical points considered in this paper.}.
The phase diagram associated with the action (\ref{double}) is shown in
Fig. 7.

\resection{Conclusion}

In the central part of this paper we applied the $S$-matrix--form factor 
approach to the regime
III of the RSOS models. These models, however, are not the only lattice models 
whose scaling limit is described by the $\varphi_{1,3}$ perturbation of $C<1$ 
conformal field theories. It is well known \cite{DF,N1,N2} that the same 
action 
(\ref{action}) corresponds to the scaling dilute $q$-state Potts model at the 
critical temperature and zero external field (with $q=4\cos^2\pi/p\in[0,4]$), 
and to the scaling $O(n)$ vector model in zero external field (with $n=2\cos
\pi/p\in[-2,2]$). The latter two models make sense for continous values 
of $q$ and $n$ through mapping onto cluster and loop models, respectively. 
Excepting special values of $p$ ($p=3$ in particular) these three models are 
characterised by different internal symmetries and then represent different
universality classes of critical behaviour. In fact, the order parameter has a 
different number of independent components in the three cases and corresponds 
to different operators (see Table 2). The fact that the three
models  are described by the same action along the renormalisation group 
trajectories specified above means that the theory (\ref{action}) admits 
different microscopic descriptions distinguished by the choice of local 
observables\footnote{Famous examples of this kind of situation are the 
equivalence between the Ising model and free neutral fermions, or that between 
the Sine-Gordon and massive Thirring models.}. 
Each description is characterised by a specific set of mutually local operators
with well defined transformation properties under the group of internal 
symmetry. Of course, the perturbing operator $\varphi_{1,3}$ appears in all 
these local sets and is invariant under the different symmetry groups. The 
observables associated to this operator, e.g. the correlation length critical
exponent $\nu=1/(2-X_{1,3})$, are the same in the three cases. 

At the conformal level, the possibility of different local descriptions appears
through the existence of different modular invariant partition functions for 
the same value of the central charge $C$ \cite{Cardy,CIZ,dFSZ}. In the S-matrix
approach away from criticality the different nature of the order parameter
leads to the existence of different scattering descriptions for the action 
(\ref{action}). They all exhibit the same spectrum and very similar analytic
form but differ from each other for the nature and the number of the 
elementary excitations (see Table 2). In this paper we used the scattering 
description based on the $Z_2$ symmetry which characterises the RSOS models. 
The massive dilute $q$-state Potts model at $T=T_c$ \cite{dilute,fractional} 
has $q+1$ degenerate vacua
located at the $q$ vertices and at the center of a hypertetrahedron living
in the $(q-1)$-dimensional space of the independent order parameter components.
The elementary excitations are the $2q$ kinks interpolating from the center 
to the vertices, and vice versa (Fig. 8). In the massive phase of the $O(n)$
model there is a single vacuum and the elementary excitations are $n$ 
ordinary particles transforming according to the vector representation of the 
group \cite{Sashapolymers,CM}. Of course, for non-integer values of $q$ and
$n$ the number of excitations is also non-integer, but this is not more 
surprising than the appearence of operators with non-integer multiplicity
in the modular invariant partition functions for the two models at criticality
(see \cite{dFSZ}). 

The different number of excitations for a given $p$ ensures that there is no
one to one correspondence between the three scattering descriptions, although
some connections certainly exists\footnote{A relation between the RSOS and 
dilute Potts scattering theory for $p=6$ was pointed out in \cite{fractional}.
The issue of the relation between the $O(n)$ and RSOS scattering descriptions
has been discussed in \cite{Smirnovcomment} from the point of view of quantum
group reduction of Sine-Gordon model.}. For each particle basis the asymptotic 
states have obvious transformation properties under the relevant symmetry group
and this fact allows a natural identification within the form factor approach 
of the interesting operators (for example the order parameter). 

As was discussed above, no matter which particle basis is used, summation 
over the intermediate asymptotic states must lead to the same result for the
correlation functions of some invariant operators, in particular the trace 
of the stress-energy tensor $\Theta(x)\sim\lambda\varphi_{1,3}(x)$. Since 
each $n$-particle contribution to the spectral sum has a distinct large 
distance behaviour $\exp(-nmr)$, the identification is expected to occur term
by term. It is easy to check comparing the results of this paper with 
those of Refs.\,\cite{dilute,CM} that this is indeed the case for the first
(two-particle) contribution to $\langle\Theta(x)\Theta(0)\rangle$.

Most of the considerations of this section can be extended to the case of 
the more general action (\ref{double}).

\vspace{1cm}
\noindent
{\bf Acknowledgements.} I thank John Cardy for interesting discussions.




\newpage

\begin{center}

\begin{tabular}{|c||c|c|c|}\hline
$\Sigma_{\Phi_1\Phi_2}^{(k)}$ & $0.5$ & $1.5$ & $2.5$ \\ \hline
$\rho_{\Theta\Theta}^{(0)}$ & $0.09$ & $0.004$ &  \\ 
                          & $(0.66)$ & $(0.55)$  &  \\ \hline
$\rho_{\Theta\Theta}^{(2)}$ & $    $ & $0.005$ & $0.0008$  \\ 
                          & $    $ & $(0.86)$ & $(0.66)$ \\ \hline
$\rho_{\Theta{\cal E}}^{(0)}$&$0.03$ & $0.007$ & $0.004$  \\ 
                          & $(0.86)$ & $(0.66)$ & $(0.55)$ \\ \hline
\end{tabular}
\end{center}
\vspace{1.5cm}
{\bf Table 1.} Accuracy $\rho_{\Phi_1\Phi_2}^{(k)}$ of the two particle 
approximation for the $k$-th moment of the correlator $\langle\Phi_1(x)
\Phi_2(0)\rangle$ for three values of the exponent (\ref{sigma}).
The numbers in parentesis are the corresponding values of $p/(p+1)$.

\begin{center}

\vspace{3cm}
\begin{tabular}{|c||c|c|c|}\hline
$   $ & RSOS   & Dilute &           \\ 
$   $ & models & Potts model & $O(n)$ model \\ \hline
$   $ &             & $q=4\cos^2\frac{\pi}{p}$ & $n=2\cos\frac{\pi}{p}$ \\  
\hline
Symmetry&$Z_2$& $ S_q $  &  $O(n)$   \\ \hline
Order parameter: & $ $ &$ $ & $$ \\ 
number of components & $1$ & $q-1$ & $n$ \\ 
scaling dimension &$X_{2,2}$& $X_{\frac{p}{2},\frac{p}{2}}$ & 
$X_{\frac{p-1}{2},\frac{p+1}{2}}$  \\ \hline
number of vacua  & $p-1$ & $q+1$ & $1$ \\ \hline
number of   & $$ & $$ & $$ \\ 
elementary excitations & $2(p-2)$ & $2q$ & $n$ \\ \hline 
\end{tabular}
\end{center}
\vspace{1.5cm}
{\bf Table 2.} Some features of the lattice models whose continuum limit is 
described by the action (\ref{action}). The notation for the scaling 
dimensions refers to (\ref{xmn}).

\newpage


\begin{figure}
\vskip 2cm
\centerline{
\psfig{figure=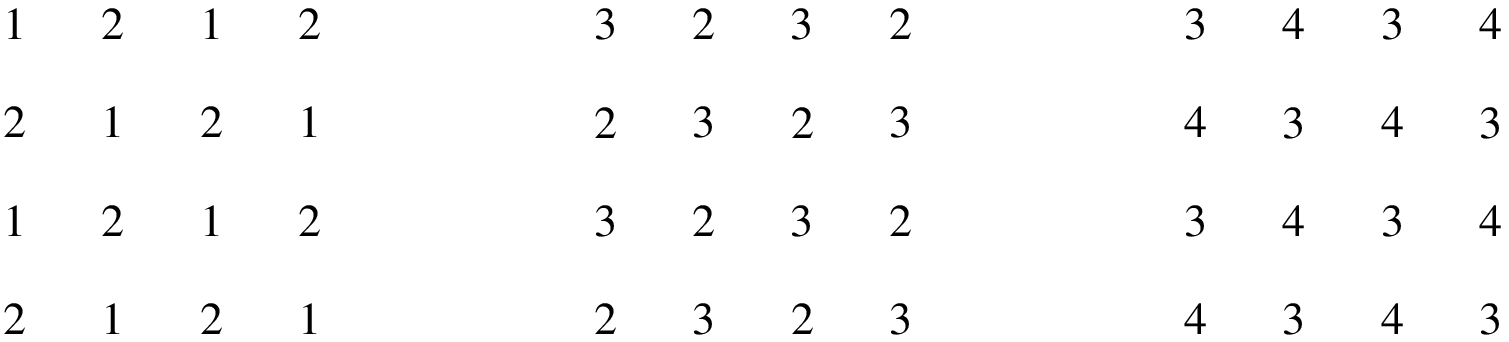}}
\vskip 1cm
{\bf Figure 1.} The ground states 1, 2 and 3 of regime III for $p=4$.
\vspace{1cm}
\end{figure}

\vspace{8cm}

\begin{figure}
\centerline{
\psfig{figure=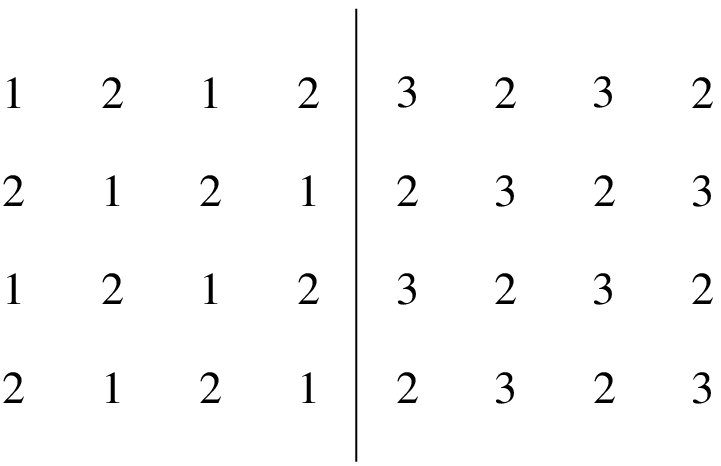}}
\vskip 1cm
{\bf Figure 2.} Domain wall (segment) between the ground states 1 and 2.
It corresponds to the space-time trajectory of the kink $K_{12}$.
\vspace{5cm}
\end{figure}
\vspace{7cm}

\newpage
\begin{figure}
\vspace{2cm}
\centerline{
\psfig{figure=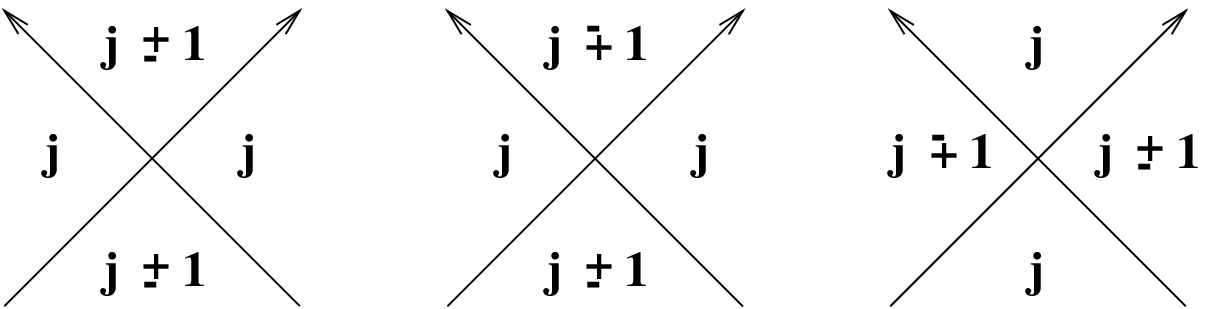}}
\vskip 1cm
{\bf Figure 3.} The two-kink scattering amplitudes $A^{\pm}_j$, $B_j$ and 
$C_j$.
\end{figure}


\begin{figure}
\centerline{
\psfig{figure=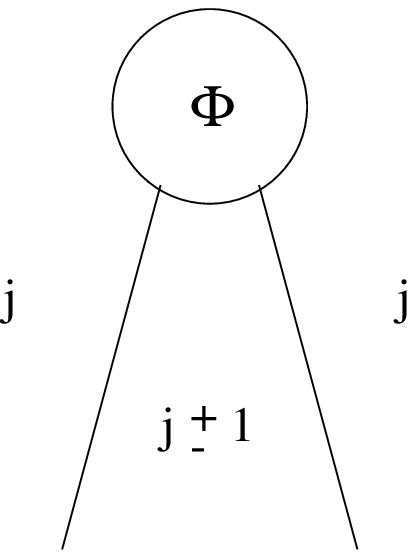}}
{\bf Figure 4.} The two-kink form factor $F^\Phi_{j,\pm}$ of the operator
$\Phi$.
\end{figure}

\newpage
\begin{figure}
\null\vskip -4cm 
\centerline{
\psfig{figure=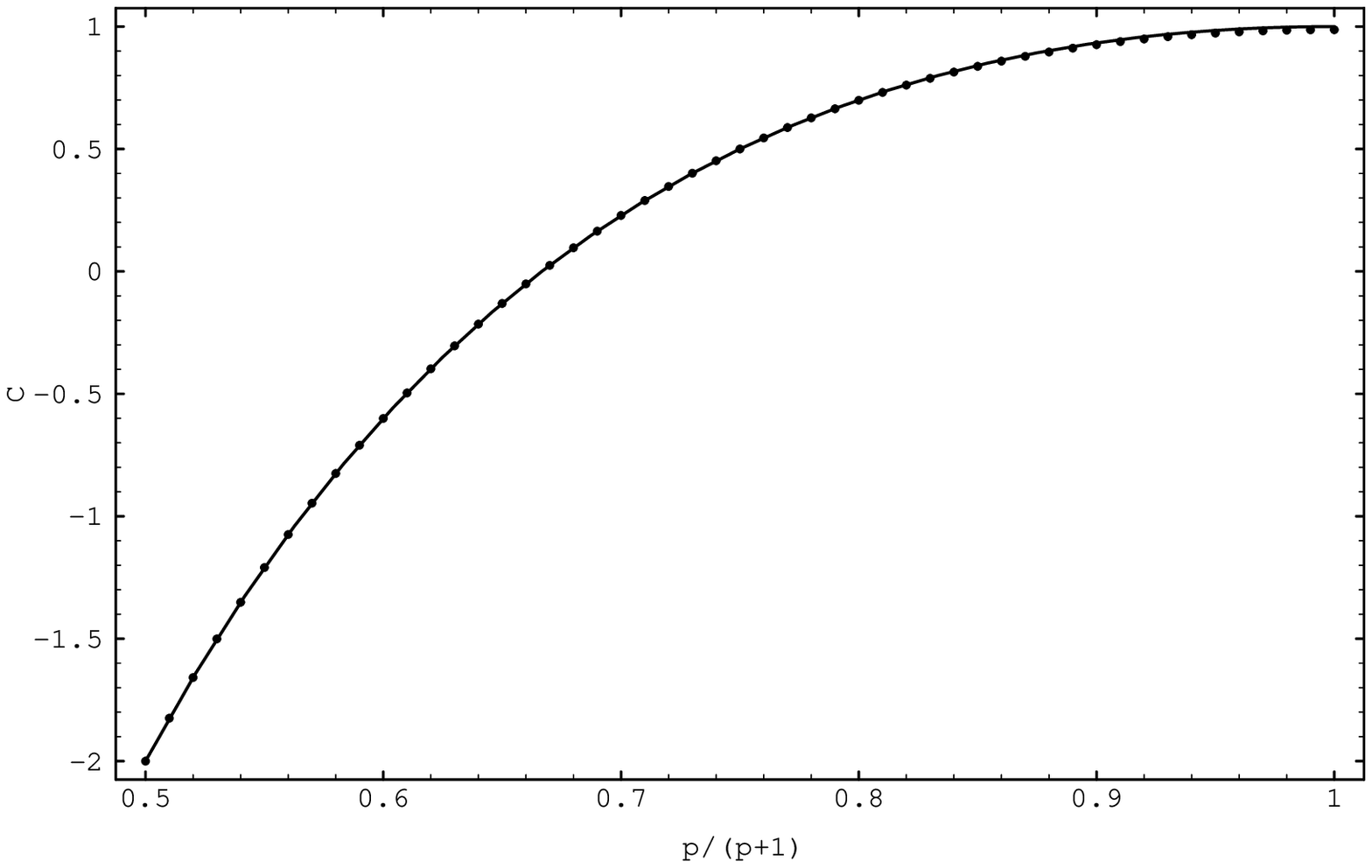}}
\vskip -5cm
{\bf Figure 5.} The two-kink approximation for the central charge sum rule
(\ref{sumc}) (dots) compared with the exact formula (\ref{c}) (continous line).
\end{figure}

\newpage
\begin{figure}
\null\vskip -4cm 
\centerline{
\psfig{figure=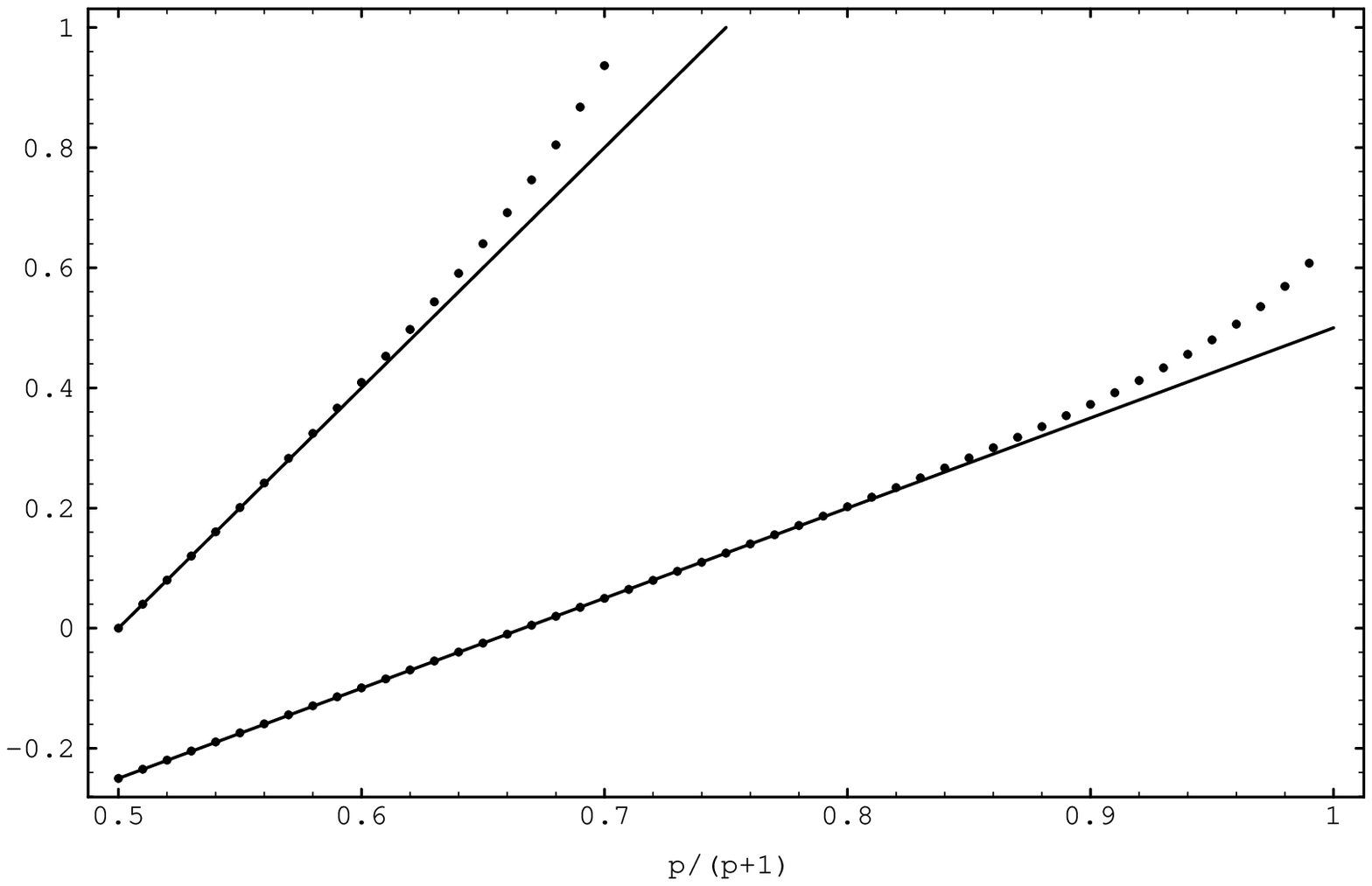}}
\vskip -5cm
{\bf Figure 6.} The two-kink approximations for the scaling dimension sum rule
(\ref{sumx}) for the operators $\Theta(x)$ (upper dotted line) and 
${\cal E}(x)$ (lower dotted line). They are compared with the exact formulae
(\ref{x13}) and (\ref{x12}), respectively (continous lines).
\end{figure}

\newpage


\begin{figure}
\vskip -1cm
\centerline{
\psfig{figure=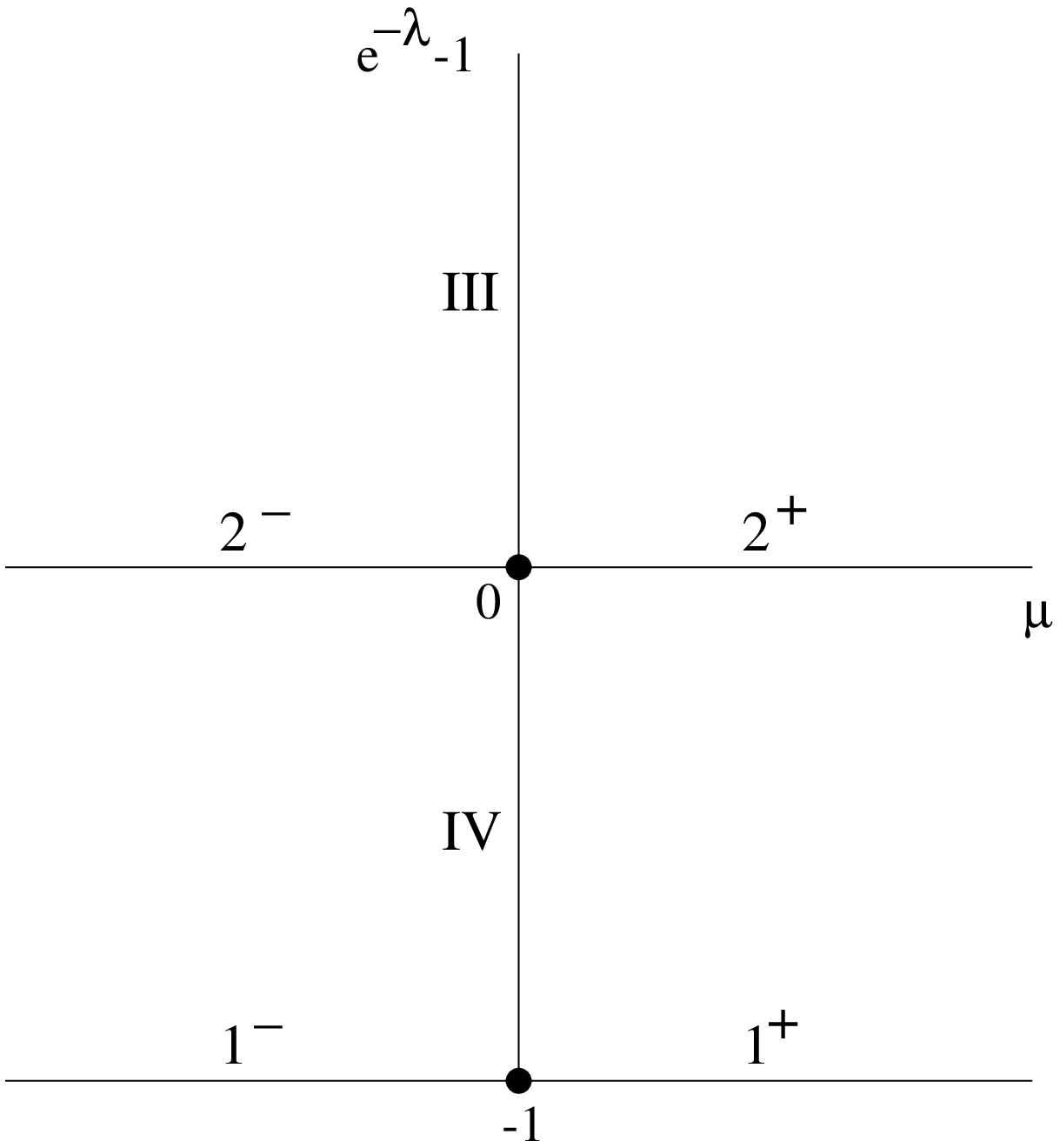}}
{\bf Figure 7.} Phase diagram associated to the action (\ref{double}) and
illustrating the different integrable branches of the RSOS models.
\end{figure}


\begin{figure}
\vskip .5cm
\centerline{
\psfig{figure=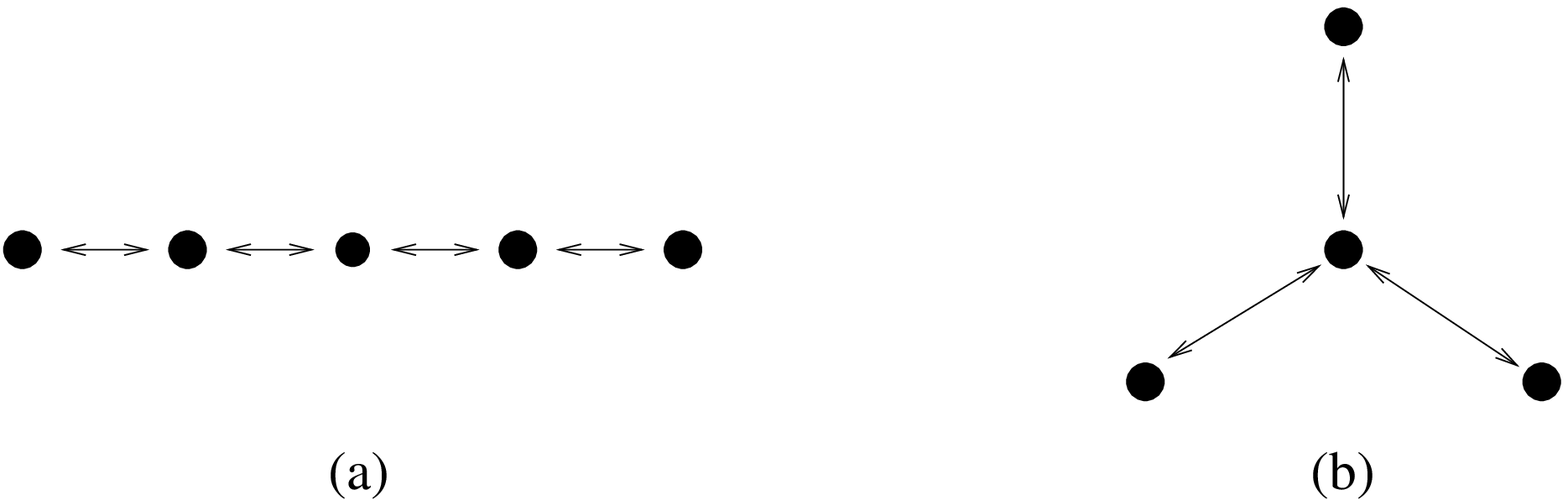}}
{\bf Figure 8.} Vacua and kinks for $p=6$ in the regime III of the RSOS models
(a), and in the massive dilute Potts model at $T=T_c$ (b).
\vspace{1cm}
\end{figure}

\end{document}